\documentclass[12pt,dvips]{article}

\newcommand{\beq}{\begin{equation}}
\newcommand{\eeq}{\end{equation}}

\usepackage{amssymb,amsmath}
\usepackage{graphicx}

\numberwithin{equation}{section}

\begin{document}

\parindent 3em
\parskip 0ex
\baselineskip 4ex
\jot 0.5cm
\tolerance=1500
\flushbottom

\begin{center}
{\LARGE \bf METALLIC SLABS:

PERTURBATIVE TREATMENTS

BASED ON JELLIUM}

\baselineskip 3ex
\vspace*{1cm}

C. FIOLHAIS$^1$, C. HENRIQUES$^{1,2}$, I. SARR\'IA$^3$, and J. M.
PITARKE$^{3,4}$

\vspace*{1cm}

$^1$Center for Computational Physics, Department of Physics, University of
Coimbra,
P-3004-516~Coimbra, Portugal

$^2$Departamento de F\'{\i}sica, Faculdade de Ci\^encias e
Tecnologia,  Universidade Nova de Lisboa, P-2825-114 Caparica, Portugal

$^3$Materia Kondentsatuaren Fisika Saila,
Zientzi Fakultatea, Euskal Herriko Unibertsitatea, 644 Posta kutxatila,
48080 Bilbo, Basque Country, Spain.

$^4$Donostia International Physics
Center (DIPC) and Centro Mixto CSIC-UPV/EHU, Donostia, Basque Country,
Spain
\end{center}

\vspace*{1.1cm}

{\Large\bf Abstract}
\vspace*{2ex}

\parbox{15cm}{\baselineskip 12pt
\small We examine first-order perturbative results based on jellium for the
surface
energy of slabs of simple metals, using various local
pseudopotentials (Ashcroft, Heine-Abarenkov and evanescent core).
The difference between the pseudopotential and the jellium potential
is averaged  along the plane parallel to the surface.
We compare these perturbative results with those of the stabilized jellium
model (a modification of the regular jellium model in which the
perturbation
appears in the energy functional right from the outset) and with the output
of other perturbative and non-perturbative
calculations.}

\baselineskip 18pt
\vspace*{1em}

\vspace*{2em}

\section{ Introduction}

\hspace*{3em}The jellium model is the simplest  model which can be
used to describe simple metals (metals with $s$ or $p$ valence
electrons). It avoids the atomic non-uniformities by replacing the ionic
cores by a positive uniform background. It describes qualitatively the
work function, but it predicts negative surface energies for metals with
high valence-electron density.

Lang and Kohn, who were the first to apply the jellium model to
surfaces~\cite{S.E.1stPP-L.K.,W.F.1stPP-L.K.},
introduced a perturbative correction to get realistic results. This
correction was simple enough to keep
most of the original simplicity: the self-consistent density was still that
of jellium and
the perturbation (difference between the lattice potential and that of the
uniforme positive background) was averaged over the surface plane. They
also improved on the description of the ion-ion interaction through the
so-called {\it{classic cleavage surface energy}}. In this
way, they obtained face-dependent
surface energies, which were always positive.

Later, the influence of
the discrete ions in the electronic density was taken into
account~\cite{S.E-2ndPP-R.D.} via
second-order  perturbation theory, which
includes the linear response of the electronic distribution to the lattice
potential~\cite{QUI-D.H.}. Rose and Dobson~\cite{S.E-2ndPP-R.D.} were the
first to work out second-order surface energy terms,
but they used the linear response of bulk jellium, in a kind of
{\it{local density approximation}}. Second-order perturbation theory
using the  linear response of a jellium slab in the {\it{Random Phase
Approximation}} has been worked out by Barnett and
coworkers~\cite{2ndPPH-Barn.} and also by Eguiluz~\cite{2ndPP.HA.-Egui.}.
These calculations, which are three-dimensional,
show a noticeable influence of the second-order term in the face-dependent
surface energies.
While the surface energies depend strongly on the exposed face in the
first-order
perturbative model of Lang and Kohn,
the second-order results
show a weaker dependence.

The stabilized-jellium model or {\it{structureless pseudopotential
model}}~\cite{SJM-P.T.S.,S.J.CAL.-FIO.,2D.PP-Mon.P.},
a modification of the regular jellium model in which the perturbation
appears in the energy functional right from the outset, includes the
perturbation
in the effective potential of the self-consistent Kohn-Sham equations and, as a
consequence, its effect in the electronic density. Originally, this
model was intended to describe flat surfaces and therefore did not include
any structure in
the averaged perturbation.
However, the dependence of
the surface energy on the atomic corrugation of a particular face of a
metal was
incorporated multiplying the flat surface results by a term based on the
liquid-drop model~\cite{SJM-P.T.S.,SJM.REV.-Per.}. The results showed a
much weaker face dependence
than in the work of Lang and Kohn, in agreement with second-order perturbative
results. Considering a dipole barrier, due to corrugation, the model was
also adapted to improve the
previous face-dependent  surface energies~\cite{SJM.F.D.-Kie.} and to
calculate face-dependent work
functions~\cite{SJM-P.T.S.,SJM.REV.-Per.,SJM.F.D.-Kie.}.

More elaborated, and therefore more computationally demanding non-perturbative
calculations, are now available for surfaces~\cite{S.E.LMTO-Skriver}. The
experimental difficulties to get
surface energies of the different surfaces to compare with the perturbative
results based on jellium were then
obviated by the predictions of these full atomistic calculations.
The results of the stabilized-jellium model were found
to be fairly realistic for several metals~\cite{comp.S.E.-Ing.}.

Some of the above-mentioned calculations were performed for slabs or thin
films,
i.e., systems made out of a few atomic layers. Slabs, which are convenient
to obtain second-order
perturbative results for
surfaces, are interesting in their own since they exhibit quantum size  and
self-compression effects.
Jellium slabs, showing quantum size effects, were examined in a seminal
paper by Schulte~\cite{QSE.-Sch}.
The second-order perturbative energies of Barnett \textit{et
al.}~\cite{2ndPPH-Barn.} and
Eguiluz~\cite{2ndPP.HA.-Egui.} were obtained for slabs. Slabs of stabilized
jellium
were recently examined by us~\cite{QSE-we}. We have shown that they are able
to describe both quantum-size and self-compression effects.

The above perturbative treatments were implemented using
Ashcroft's empty-core pseudopotential~\cite{PP-Ash} or the local
Heine-Abarenkov pseudopotential~\cite{PP.H-A.}.
In this paper we study aluminum slabs, with 9 and 17 layers, along the
lines of Lang-Kohn's perturbation theory
using the Ashcroft, the Heine-Abarenkov, and the recent evanescent-core
pseudopotential~\cite{PP.Evan.}, which has the advantage of having a smooth
repulsion.
We compare our results
with those of stabilized jellium and with other perturbative and
non-perturbative calculations.
Finally, we  refer to the possibility of considering
stabilized jellium as a zero-order description of a metal surface.

The ultimate goal of our research is to perform systematic first and
second-order perturbative calculations
for surfaces of simple metals with the  evanescent-core pseudopotential.
This will be an extension to surfaces
of the systematic treatment of the energetics and mechanical properties  we
have  made for metallic solids in different
crystal structures~\cite{EC.bulk-Fern.}.
Although \textit{ab initio} calculations are nowadays clearly the method of
choice for bulk or surface
systems, perturbative treatments still have their role to describe trends
along the the periodic table
and along different crystallographic structures. Above all, they can provide
understanding of the
physics of metal cohesion.

The outline of this paper is as follows. In Section 2, the theoretical
background
is provided, in Section 3 the perturbative results are presented.
Conclusions appear in Section 4.

\vspace*{1em}

\section{ Perturbative corrections to jellium model}

\subsection{{\it\textbf{Electronic and Madelung
subsystems}}}

\hspace*{3em}In order to study a neutral metallic system formed by a
fixed lattice of ions and
electrons with density $n(\vec{r})$, we start with  a superposition of two
simple systems: the
valence electrons moving in a positive background of density $n_+(\vec{r})
= \bar n$ inside the
metal ($n_+(\vec{r})=0$ outside), where $\bar{n}$ is the mean electronic
density, and the ions lattice
embedded in a negative background with the same density and size as the
positive one.

The first of these subsystems, referred to as the electronic subsystem, is
the jellium
model. We denote its ground-state energy by $E_J[n_0]$, $n_0$ being its
electronic density which we
assume to be a reasonable approximation to $n({\vec{r}})$.

The second subsystem, called {\it Madelung subsystem}, has energy $E_M$.
If one wishes to correct the jellium description perturbatively, one should
consider the interactions within the Madelung system and between this and the
electronic subsystem, $E_{M,e}$. The total energy is
\begin{equation}\label{ET}
E = E_J[n_0] + E_M + E_{M,e}.
\end{equation}

The energy of the jellium model (electronic subsystem) is, in
atomic units, given by the following density functional:
\begin{multline} \label{Funcgeleia}
E_J[n_0] = T[n_0] + E_{xc}[n_0] + \frac{1}{2} \int{d^3r\int{d^3r'\frac{
n_0(\vec{r'})n_0(\vec{r'})}{\left|\vec{r} -\vec{r'}\right|}}} + \\
+ \frac{1}{2}
  \int{d^3r\int{d^3r'\frac{n_+(\vec{r'})n_+(\vec{r'})}{\left|\vec{r}
-\vec{r'}\right|}}} -
\int{d^3r\int{d^3r'\frac{n_0(\vec{r'})n_+(\vec{r'})}{\left|\vec{r}
-\vec{r'}\right|}}},
\end{multline}
where $T[n_0]$ is the noninteracting kinetic energy, $E_{xc}[n_0]$ the
exchange-correlation energy (often evaluated in the local density
approximation,
which is incidentally adequate for jellium surfaces ~\cite{Pit, XC-K.P.}),
and the other terms represent, respectively, the electronic repulsion, the
self-repulsion of the positive background and the attraction between the
electrons and the positive background.

On the other hand, the Madelung energy is given by
\begin{equation} \label{Madelung}
E_M=\frac{1}{2}\int d^3r\int
d^3r'\frac{n_+(\vec{r})n_+(\vec{r'})}{\left|\vec{r}
-\vec{r'}\right|}  +
\frac{1}{2}\sum_l \sum_{l'\neq l} \frac{Z^2}{\left|\vec{R}(l)-\vec{R}
(l')\right|}
-\int d^3r\sum_{l} \frac{Zn_+(\vec{r})}{\left|\vec{R}(l)-\vec{r}\right|},
\end{equation}
where the summations run over the ionic positions $\vec{R}(l)$ and $Z$ is
the charge of each ion. The first term is
the self-repulsion of the negative background,
the second the Coulomb repulsion of the ions
and the third the interaction between the negative background and the ions.

\subsection{\noindent{{\it \textbf {First-order perturbative
correction}}}}

\hspace*{3em}In order to simplify the calculations, it is convenient to
use a local pseudopotential, $v_{ps}$, to
represent the ion-electron interactions (from now on, by electrons we mean
valence-electrons).
The pseudopotential, due to a pseudo-ion of charge $Z$ located at the
position $\vec{R}(l)$, can be
written as a sum of two contributions: an attractive long-range Coulomb
part and a short-range repulsive part:
\begin{equation}\label{pseudo}
v_{ps}(|\vec{r} - \vec{R}(l)|) = -\frac{Z}{|\vec{r} - \vec{R}(l)|} +
\omega_{R}(|\vec{r} - \vec{R}(l)|).
\end{equation}

The energy arising from the interactions between the Madelung  and the
electronic subsystems can be written in first order as:
\begin{multline}\label{Intersubsis}
E_{M,e}=  E^{(1)}_{ps}[n_0] + E_{M,+J} = \\
= \left\{\int{d^3r\sum_l{{n_0(\vec{r})\left[v_{ps}\left({\vec{r}-\vec{R}
(l)}\right)\right]}}} +
\int{d^3r\int{d^3r'\frac{n_0(\vec{r})n_+(\vec{r'})}{\left|\vec{r}-\vec
{r'}\right
|}}}\right\} + \\
+  \left\{ \int{d^3r\sum_l{\frac{Z n_+ (\vec{r})}{\left|{\vec{R}(l) -
\vec{r}}\right|}}}
- \int{d^3r\int{d^3r'\frac{n_+(\vec{r})n_+(\vec{r'})}{\left|\vec{r}-\vec
{r'}\right|}}}\right\}.
\end{multline}

The terms in the first brackets of (\ref{Intersubsis}) represent
the interaction between the Madelung subsystem and the electrons of the
electronic subsystem. They correct the jellium-electron interactions
(last term in the right-hand side of (\ref{Funcgeleia})).
The difference ${\delta}v(\vec{r})$ between the potential of a pseudo-ions
lattice and the potential of the jellium
background appears in the first-order term,  which is the first correction
to the jellium model:
\begin{equation}\label{FirstPseudo}
E_{ps}^{(1)}[n_0] = \int{d^3r \,{\delta} v(\vec{r}) \, {n_0(\vec{r})}},
\end{equation}
where
\begin{equation}\label{deltav}
{\delta}v(\vec{r}) = \sum_l{v_{ps}\left({\vec{r}-\vec{R}(l)}\right)} +
\int{d^3r'\frac{n_+(\vec{r'})}{\left|\vec{r}-\vec{r'}\right|}}.
\end{equation}

The second brackets of (\ref{Intersubsis}) include terms representing
the interaction between the
Madelung  subsystem and the positive background of the electronic
subsystem. Taking advantage of the
pseudopotential form (\ref{pseudo}) we may rewrite it as
\begin{equation}\label{MJcRep}
E_{M,+J} = - \int{d^3r \, {\delta} v(\vec{r}) \, {n_+(\vec{r})}} +
\int{d^3r\sum_l{\omega_{R}\left({\vec{r}-\vec{R}(l) }\right)}{n_+(\vec{r})}}.
\end{equation}

If we add $E_M$ of (\ref{Madelung}) to $E_{M+J}$ (as written in the
second brackets of
(\ref{Intersubsis})), we find the second  correction to the jellium
model, the one which improves the description of the ion-ion interaction:
\begin{equation}\label{Corrion}
E_M + E_{M,+J} =
\frac{1}{2}\sum_{{l,l',l{\neq}l'}}{{\frac{Z^2}{\left|\vec{R}(l)-\vec{R
}(l')\right|}}}
-\frac{1}{2}\int{d^3r\int{d^3r'\frac{n_+(\vec{r})n_+(\vec{r'})}{\left|\vec{r}
-\vec{r'}\right|}}},
\end{equation}
which amounts to  replacing the self-repulsion of the background by the
repulsion of the point ions.

The described perturbative approach is based on general
density functionals and may be applied as well to solids, surfaces, slabs
or clusters.
Lang and Kohn~\cite{S.E.1stPP-L.K., W.F.1stPP-L.K.} used it to
calculate surface properties of metals. However, they replaced the
difference potential
$\delta v(\vec{r})$ in (\ref{FirstPseudo}) by its average along the
plane parallel to the surface, ${\delta}v(z)$ (z being the direction
perpendicular to the surface),
keeping therefore the one-dimensionality of the underlying jellium model,
for which
$n_0(\vec{r})\equiv{n_0(z)}$. They wrote:
\begin{equation}\label{meanplan}
{\delta}v(z) = \langle \delta v(\vec{r})\rangle=
\sum_k \hbox{ }{\bar{n}}\hbox{ }2 \pi d \int dx_{||}{\hbox{
}v_{ps}\left( \left[
x_{||}^2 + |z-R_z(k)|^2
\right]^{-1/2}\right)} x_{||}-\varphi_+(z),
\end{equation}
where $k$ runs over the ionic planes,  $d$ denotes the inter-layer spacing,
$x_{||}$ is the
distance in the plane parallel to the surface from any ion, $R_z(k)$ is the
position of the kth
plane of ions, and $\varphi_+(z)$ is the potential due to the uniform
positive background.

Lang and Kohn used the Ashcroft empty-core pseudopotential~\cite{PP-Ash},
which  requires for each
metal a single parameter, the core radius $r_C$:
\begin{equation}\label{Ash}
v_{ps}^{A} \left(|\vec{r} - \vec{R}(l)|\right) =
\begin{cases}
      0, & \text{    $|\vec{r} - \vec{R}(l)| < r_c$} \\
     - Z/{|\vec{r} - \vec{R}(l)|},      & \text{     $|\vec{r} -
\vec{R}(l)| \geq r_c$}.
\end{cases}
\end{equation}

A better  pseudopotential is the
local form of the Heine-Abarenkov pseudopotential~\cite{PP.H-A.}, which
includes two parameters, $R_C$ and $u$, the first measuring the core radius
and the second the amount
of repulsion in the core,
\begin{equation}\label{HeineAb}
v_{ps}^{HA} \left(|\vec{r} - \vec{R}(l)|\right)
=
\begin{cases}
      Zu/{R_c}, & \text{    $|\vec{r} - \vec{R}(l)| < R_c$} \\
     {- Z/|\vec{r} - \vec{R}(l)|},      & \text{    $|\vec{r} -
\vec{R}(l)| \geq R_c$}.
\end{cases}
\end{equation}

A recent pseudopotential incorporating an exponential decay of the core
repulsion, devised for evaluating systematically the energetics of simple
metals, is
the evanescent core pseudopotential~\cite{PP.Evan.},  which also depends on two
parameters, $R$ and $\alpha$:
\begin{multline}\label{Evan}
v_{ps}^{ec} \left(|\vec{r} - \vec{R}(l)|\right) = -\frac{Z}{|\vec{r} -
\vec{R}(l)|}+ \\
+\frac{ZA}{R}
e^{-{\alpha}\frac{|\vec{r} - \vec{R}(l)|}{R}}+
\frac{Z}{|\vec{r} - \vec{R}(l)|}\left({1+B \frac{|\vec{r} -
\vec{R}(l)|}{R}}\right)
e^{-{\alpha}\frac{|\vec{r} - \vec{R}(l)|}{R}},
\end{multline}
where $A$ and $B$ are simple functions of $\alpha$.
The smoothness of this potential assures good convergence of its Fourier
transform and its suitability
to second-order perturbative calculations.
It yields overall good results for simple metal solids and
clusters~\cite{PP.Evan.,EC.bulk-Fern., EC.Clusters,
EC.surf.clust}, when the parameters $R$ and $\alpha$ are fitted to
solid-state information.

Perturbative first-order theory can be applied to slabs along the lines
of Lang and
Kohn using these or other local pseudopotentials. It can also be done
exactly, i.e., without taking any
average~\cite{2ndPPH-Barn.,2ndPP.HA.-Egui.}, but then part of
the simplicity of the jellium model is  lost.

If one is working with slabs, the surface energy can be extracted from the
total slab
energy per unit area by subtracting the corresponding bulk energy:
\begin{equation}\label{se}
\sigma (L) = {\frac{1}{2A}}\left[E(L)-{\bar n} \,L \,A
\,\epsilon^{bulk}\right],
\end{equation}
where $A$ is an area, $L$ is the slab width (the width of the jellium
background in our system) and
$\epsilon^{bulk}$ is the total energy per particle in the bulk. We can
decompose the slab surface energy in various parts. For the
functional of (\ref{ET}) and considering (\ref{Intersubsis}), the
surface energy reads as:
\begin{equation}\label{sigdec}
\sigma = \sigma_J + \sigma_M + \sigma_{ps}^{(1)}  + \sigma_{M,+J}.
\end{equation}
The first term in the right-hand side is the jellium surface energy. The second
is the Madelung surface energy, for which the following classical cleavage
formulae may be used, as Lang and Kohn did~\cite{S.E.1stPP-L.K.}
\begin{equation}
\sigma_M\approx\alpha Z\bar{n},
\label{eq16}
\end{equation}
where $\alpha$ is a tabulated constant for each face of a given crystal
structure.
The third term is the first-order pseudopotential surface energy,
which may be approximated  using the average value for the perturbative
potential given by (\ref{meanplan}),
\begin{equation}
\sigma_{ps}^{(1)} \simeq \frac{1}{2} \int_{-\infty}^{\infty}dz\ \delta
v(z)[n(z)-n_+(\vec{r})].
\end{equation}

Finally, $\sigma_{M,+J}$ is a cleavage piece which is
different from zero only when the ion cores appear out of the jellium
surface. We may use,
for this term, the expression of \cite{2D.PP-Mon.P.}, where the
potential is averaged
in planes parallel to the surface, as previously done for the difference
potential:
\begin{equation}
\sigma_{M,+J}\simeq-\frac{1}{2}\int_{-\infty}^\infty dz
\left<\sum_l\omega_R\left(\vec{r}-\vec{R}(l)\right)\right> \bar{n},
\end{equation}
where the angular brackets denote the surface average.

When the width of the slab approaches infinity, the surface energy of the
slab  approaches the surface
energy of the semi-infinite system.

\subsection{\noindent{{\it\textbf{Stabilized jellium}}}}

\hspace*{3em}In the stabilized-jellium model, which  is based on the
perturbative-variational concept
of Monnier and Perdew~\cite{2D.PP-Mon.P.}, the perturbative potential,
conveniently averaged, is included in the
Kohn-Sham equations. The new self-consistent density is  a better approach
to the
real density than the jellium one.  The corrections
to the jellium description of the ion-ion and electron-ion interactions are
now averaged out: the
self-repulsion of the jellium positive background inside each Wigner-Seitz
sphere $\tilde\epsilon$ is subtracted
(the jellium is supposed to describe well the repulsion between cells) and,
in the perturbative energy,
  $\delta{v}(\vec{r})$ is taken to be constant inside the metal
and equal to its average over the volume of the Wigner-Seitz
sphere, ${\langle\delta v\rangle}_{WS}$.
The stabilized jellium energy functional is given by
\begin{displaymath}
E_{SJ}[n]=E_J[n]-\tilde\epsilon\int d^3r\ n_+(\vec{r})+\langle\delta
v\rangle_{WS}\int d^3r\ \theta(\vec{r})n(\vec{r})
\end{displaymath}
or, using the equality~\cite{SJM-P.T.S.} $\tilde\epsilon=\langle\delta
v\rangle_{WS}-e_M-\bar\omega_R$,
where $e_{M}$ is the (bulk) Madelung energy per particle and
$\bar{w}_{R}$ is the pseudopotential repulsion averaged in the Wigner-Seitz
sphere,
\begin{eqnarray}\label{efu}
E_{SJ}\lbrack{n}\rbrack=E_{J}\lbrack{n}\rbrack+
\left(e_{M}+\bar{w}_{R}\right)\int d^3{r}\,{n_+(\vec{r})}
+{\langle\delta v\rangle}_{WS}\int d^3{r}\,\frac{n_+(\vec{r})}{\bar{n}}
\,\left[n(\vec{r})-n_+(\vec{r})\right].
\end{eqnarray}

While in the jellium model the energy per particle of the bulk system has a
single minimum at a density close to
that of sodium
\begin{displaymath}
\frac{d \epsilon_J^{bulk}}{d n}=0
\end{displaymath}
in the stabilized-jellium model that energy has a minimum for each
metal at the corresponding experimental
density
\begin{displaymath}
\frac{d \epsilon_{SJ}^{bulk}}{d n}\left|_{n=n_{exp}}\right.=0.
\end{displaymath}
This condition is fulfilled by adjusting a pseudopotential parameter.

The second term in the right-hand side of (\ref{efu}) does not
contribute to the surface energy.
The stabilized jellium functional has been recently applied to metal
slabs~\cite{QSE-we}, the conclusion being that the
surface energy obtained from (\ref{se}) and (\ref{efu})
\begin{equation}\label{sig}
\sigma_{SJ} = \sigma_{SJ,J} + \sigma_{SJ,ps}
\end{equation}
gives a reasonable description of aluminum slabs in comparison
with {\it ab initio} results, but it fails for lithium slabs, a case
where the necessity for a non-local pseudopotential {is  known and for
which} first-principles calculations { showed untypical features for a
simple metal}~\cite{Li}. Note that the first  term of (20) is similar to
the jellium surface energy but is evaluated with the
self-consistent density obtained from the functional (\ref{efu}). On the
other hand,
the second term differs from the Lang-Kohn perturbative term by the use of
a 3D average for the perturbative potential instead of a
2D one
and by the use of the self-consistent stabilized jellium electron density
instead of the jellium electron density.
In summary: although inspired by it, the stabilized jellium model is
not perturbation theory.

In \cite{QSE-we}, we have shown that the application of the stabilized
jellium model to slabs leads
to quantum size effects, i.e., fluctuations in the surface energy and work
function,
which are similar to those known for the jellium model but are around
more realistic
values. Moreover, we have shown that, fixing the width of the slab,  energy
minimization
with respect to background variation
leads to a higher background density inside the slab, i.e., the system
tends to self-compress.

Using functional~(\ref{efu}) no difference shows up between different
crystallographic faces.
However, \textit{ad-hoc} modifications of stabilized jellium have been
proposed to describe the difference between various exposed
faces~\cite{SJM-P.T.S.,SJM.REV.-Per.,SJM.F.D.-Kie.}. In these approaches,
the self-consistent density is obtained by considering a face-dependent
 potential  but, for the sake of realism,  the use of the latter is avoided
in the
final evaluation of the first-order surface energy. This methodology describes
reasonably well the face-dependence found by more sophisticated methods.

A modification of stabilized jellium has been made by Montag, Reinhard, and
Meyer~\cite{SAJM-Montag},
who tried to incorporate the cleavage energy in a phenomenological
way (by fitting to empirical surface energies).

\vspace*{1em}

\section{ Results}

\hspace*{3em}We have considered jellium slabs corresponding to 7 and 19
layers of  aluminum (fcc lattice),
cut along the  three main planes (111), (100)
and (110) (by decreasing order of planar density and, therefore, by increasing
order of interplanar distance). We expect
perturbation theory based on jellium to converge better for the
planes which are most close-packed and therefore more similar to a flat
surface.

Fig. 1 represents our three pseudopotentials for aluminum. The
following values for the pseudopotentials parameters have
been used (all in atomic units, except $\alpha$, which is dimensionless):
$r_c$=1.12 (Ashcroft); $R_c$=1.4017 and
$u$=-0.3921 (Heine-Abarenkov); and $R$=0.317, $\alpha$=3.512 (evanescent core).
The first value is simply derived from a stability condition
for the bulk energy within first-order perturbation theory.
The second pair of values come from a stability condition for the bulk
energy at the experimental density and from
matching the  bulk modulus (within second-order perturbation theory) to the
experimental
value~\cite{2ndPP.HA.-Egui.}. The last pair of values arise again from a
stability condition for the bulk energy
within second-order and from
matching first-order values~\cite{PP.Evan.} to all-electron values
for the number of electrons in the interstitial region (zone
between the Wigner-Seitz  cell and the inscribed sphere). It has been shown
that the latter requirement does not differ much from the demand for a
realistic bulk modulus~\cite{EC.bulk-Fern.}.

Jellium surface energies $\sigma_J$ are straightforward to evaluate.
The cleavage energy $\sigma_M$ has been
taken from~\cite{2D.PP-Mon.P.}. We note that the differences between
the values of $\sigma_M$ for various faces are large.
The term $\sigma_{M,+J}$ is small: it may be even zero depending on the
size of the core
radius~\cite{2D.PP-Mon.P.}. The perturbative potentials $\delta v(z)$
corresponding to the different pseudopotentials are represented in Fig. 2. They
enter in the
calculation of $\sigma_{ps}^{(1)}$. We note the better smoothness of
the $\delta v (z)$ arising from the evanescent core potential.

Table 1 shows total surface energies, together with their components,
obtained with the three pseudopotentials for the slabs with 7 and 19
layers. The table illustrates the importance of the positive Madelung
contribution. For the same pseudopotential the
surface energy increases when
going from the  (111) face to the  (110) face. The results show a
strong dependence on the
pseudopotential, with the Ashcroft result, which in principle is the most
unrealistic, being discrepant from the other two (it is always
bigger). Without the Madelung term, the  perturbative correction of
Lang and Kohn
would lead to a positive surface energy only for the (111) face,
precisely that considered by those authors.

Table 2 shows the { flat} stabilized jellium results for aluminum slabs with {
thicknesses corresponding to} 7 and 19 layers. The total surface
energies  are in general very different not only from the
jellium ones but also from first-order perturbation results.

{In spite of the quantum size oscillations,} slabs may be used to estimate
the surface
energy of the semi-infinite system. Indeed, the slab with
thickness corresponding to 19 layers is a good
approximation to the latter. Table 3 allows for comparing our
previous slab results with semi-infinite results (Lang-Kohn,
stabilized jellium, and  other perturbative
and non-perturbative results). Actually, the non-perturbative result
was obtained with the plane-wave pseudopotential method for 12
layers separated by a vacuum of 6 layers, but this should represent
well the semi-infinite system.
Accepting this non-perturbative  as the best result, we have to conclude that
its agreement with
the (111) surface energy of Lang and Kohn, who used the Aschroft potential,
is accidental.
The same pseudopotential for any other face leads to disagreement.
However, the most striking conclusion of Table 3 is that the
face-dependent stabilized jellium
model can emulate quite well the second-order perturbative result of
Rose and Dobson and  also the non-perturbative calculation, with the
single exception of the (110) case.

\vspace*{1em}

\section{ Conclusions}

\hspace*{3em}We have studied within perturbation theory surface energies
of metallic slabs taking as zero-order the
jellium model. Our results allow us to conclude that
first-order surface energies depend strongly on the pseudopotential
used. For aluminum slabs,
the Heine-Abarenkov and the evanescent
core potentials give similar results, while the Ashcroft potential differs
from those two
(it differs more  for the least dense surfaces).
It should therefore be used with some caution.
On the other hand, the first-order perturbation turns out to be too large
for the least
dense surfaces, being imperative to
correct it through second-order terms. We are implementing second-order
perturbative
calculations using the slab response function. We pointed out the
importance of the Madelung energy, which should therefore be
evaluated as far as possible without any approximations.

We stressed the usefulness of the stabilized jellium model, which keeps
the essential simplicity of jellium, while  reproducing various
first principles results for slabs and surfaces.
As an extension of that model, we may take as  zero's order the
stabilized jellium energy instead
of the jellium one (since the perturbation is strong for all aluminum
faces, it should,
after all, be a better starting point for
perturbation theory) and to take as first-order perturbation, without
averaging,
the difference between the pseudopotential of stabilized jellium and an
adequate
pseudopotential. Work along these lines is in progress.

\vspace*{3em}

\noindent{\Large\bf{Acknowledgments}}

\vspace*{1em}

The authors gratefully acknowledge J. P. Perdew (Tulane University, USA)
for useful discussions. They also acknowledge F. Nogueira (Coimbra
University, Portugal) for providing some bulk results
and for helping in preparing the manuscript. This project has been
supported by  the Portuguese PRAXIS XXI
Program (Project PRAXIS/2/2.1/FIS/473/94) and by the University
of the Basque Country, the Basque Hezkuntza, Unibertsitate eta Ikerketa Saila,
and the Spanish Ministerio de Educaci\'on y Cultura.

\newpage

{\centerline{\Large\bf Figure captions}}

\vspace*{3em}

Fig. 1:

Three pseudopotentials for aluminum used in this work as function of
radial distance.

\vspace*{4cm}

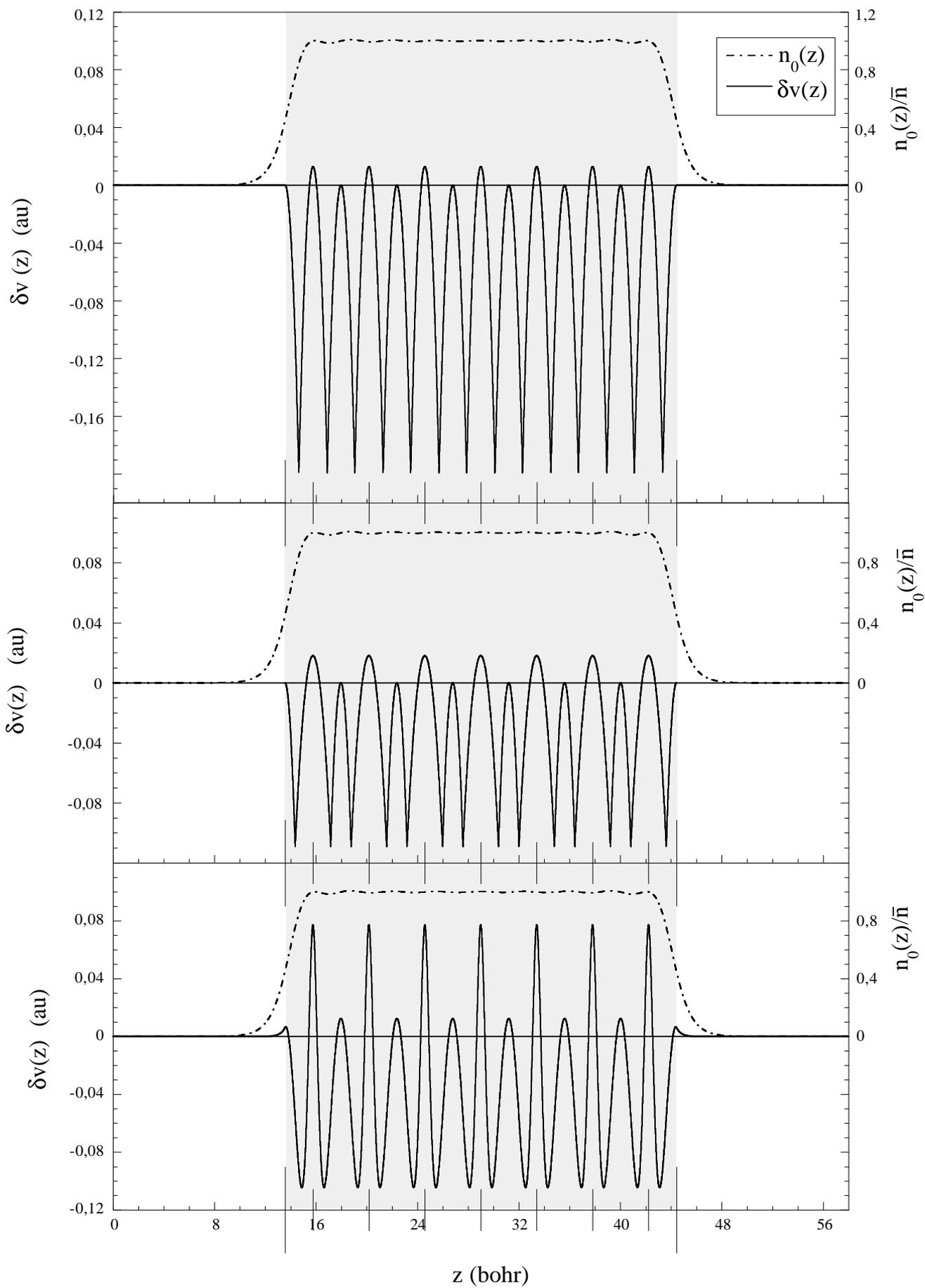
Fig. 2:

Jellium electronic density $n_0 (z)$ and perturbative potential $\delta v(z)$
contributing to first-order
energy of slab with 7 layers (111 face). Three pictures refer, from
top to bottom, to Ashcroft, Heine-Abarenkov and evanescent core
pseudopotentials. The shaded area indicates the background jellium.

\begin{table}[H]
\caption{First-order surface energies for slabs with 7 and 19 layers of
aluminum cut along, respectively, the (111), (100) and the (110) planes.
$\sigma_J$,
$\sigma_M$, $\sigma_{ps}^{(1)}$, and $\sigma_{M,+J}$ are the terms of the
surface
energy, $\sigma$, following (2.15). The
Perdew-Wang~\cite{XC-P.W.} correlation functional was used in the jellium term.
The other terms were calculated using the approximations expressed by (2.16),
(2.17), and (2.18). All values are in erg/cm$^2$.}
\vspace{1.5cm}
\begin{center}
\begin{tabular}{|c|c|c|c|c|c|c|} \hline
             & $\sigma_{J}$&   $\sigma_{M}$           &                 &
$\sigma_{ps}^{(1)}$ &      $\sigma_{M,+J}$         & $\sigma$ \\
     Face &   7 layers  &  & Pseudopotential &     7 layers
& & 7 layers \\
             & 19 layers   &              &                 &    19 layers
&               & 19 layers \\
\hline
\hline

             &    -610.4   &     408.6    &                 &       1050.5
&       0       & 851.7 \\
             &    -602.7   &     408.6    &     Ashcroft    &       1053.5
&       0       & 856.4 \\
\cline{2-7}

             &   -610.4    &     408.6    &                 &        680.9
&       0       & 479.1 \\
     Al(111) &   -602.7    &     408.6    & Heine-Abarenkov &        677.8
&       0       & 483.7\\
\cline{2-7}

             &   -610.4    &     408.6    &                 &        544.9
&     -92.7     & 250.4 \\
             &   -602.7    &     408.6    &    Evanescent core  &        543.5
&     -92.7     & 256.7\\
\hline

\hline

             &   -613.8    &    1802.8    &                 &        405.3
&       0       & 1594.3 \\
             &   -603.4    &    1802.8    &     Ashcroft    &        410.8
&       0       & 1610.2\\
\cline{2-7}

             &   -613.8    &    1802.8    &                 &       -108.3
&       0       & 1080.7 \\
     Al(100) &   -603.4    &    1802.8    & Heine-Abarenkov &       -107.7
&       0       & 1091.6\\
\cline{2-7}

             &   -613.8    &    1802.8    &                 &        -168.1
&     -182.7    & 838.1 \\
             &   -603.4    &    1802.8    &     Evanescent core &        -168.2
&     -182.7    & 848.5\\
\hline

\hline

             &   -585.6    &     5540.3   &                 &        -1588.6
&       0       & 3366.1 \\
             &   -608.3    &     5540.3   &    Ashcroft     &        -1581.7
&       0       & 3350.3\\
\cline{2-7}

             &   -585.6    &      5540.3  &                 &        -2315.2
&     -15.4     & 2624.1 \\
     Al(110) &   -608.3    &      5540.3  & Heine-Abarenkov &        -2306.2
&     -15.4     & 2610.4\\
\cline{2-7}

             &   -585.6    &     5540.3   &                 &        -1986.8
&     -588.7    & 2379.2 \\
             &   -608.3    &     5540.3   &    Evanescent core  &
-1978.1
&     -588.7    & 2365.1 \\
\hline
\end{tabular}
\end{center}
\label{tab1}
\end{table}

\begin{table}[H]
\caption{{ Flat} stabilized jellium surface energies (no corrugation
included in
the calculations) for slabs with { thicknesses corresponding to} 7
and 19 layers.
$\sigma_{SJ,J}$ and
$\sigma_{SJ,ps}$ are the terms of the stabilized jellium surface
energy, $\sigma_{SJ}$, following (\ref{sig}). The
Perdew-Wang~\cite{XC-P.W.} correlation functional was used in the
jellium term. All values are in
erg/cm$^2$.}
\vspace{1.5cm}
\begin{center}
\begin{tabular}{|c|c|c|c|} \hline
     &            &               &       \\
   thicknesss   &  $\sigma_{SJ,J}$ &  $\sigma_{SJ,ps} $ &  $\sigma_{SJ}$ \\
  &            &               &       \\
\hline
\hline
7 layers of Al(111)   &  -447.9           &    1369.8    & 921.9       \\
19 layers of Al(111)  & -450.6            &    1376.0    & 925.4       \\
\hline
7 layers of Al(100)   & -445.9            &     1363.2   & 917.3       \\
19 layers of Al(100) & -450.2     &     1376.2     & 926.0       \\
\hline
7 layers of Al(110)   & -460.1      &  1405.1   & 945.0       \\
19 layers of Al(110)&   -451.5    & 1370.8           & 919.3       \\
\hline
\end{tabular}
\end{center}
\label{tab3}
\end{table}

\begin{table}
\caption{Surface energies obtained by first-order
perturbative theory (Lang-Kohn approach ~\cite{S.E.1stPP-L.K.} with
the Ashcroft
pseudopotential), in the second column,
and by other methods.
The third column  refers to flat stabilized jellium~\cite{SJM-P.T.S.},
$\sigma_{SJ}^{flat}$, the fourth to
face-dependent stabilized jellium~\cite{SJM.F.D.-Kie.}, $\sigma
_{SJ}^{face}$, the fifth to the
perturbative-variational  method of Monnier and Perdew~\cite{2D.PP-Mon.P.},
$\sigma_{MP}$, the sixth to the second-order
perturbative theory of Rose and Dobson~\cite{S.E-2ndPP-R.D.},
$\sigma_{RD}$, and the seventh to Sch\"ochlin, Bohnen, and Ho~\cite{ATOM}
first-principles non-perturbative results,
$\sigma_{SBH}$. In these works the semi-infinite system is considered except
in the last one, where a slab with 12 layers was taken. All calculations
were done in
the Local Density Approximation.
All values are in erg/cm$^2$.}
\vspace{1.5cm}
\begin{center}
\begin{tabular}{|c|c|c|c|c|c|c|}
\hline
  & & & & & &\\
Surface & $\sigma$ & $\sigma_{SJ}^{flat}$ & $\sigma_{SJ}^{face}$  &
$\sigma_{MP}$ & $\sigma_{RD}$
& $\sigma_{SBH}$  \\
  & & & & & &\\
\hline
\hline
   Al(111)&  842  &  925      &        938       &    643        &     1065
&      939       \\
\hline
   Al(100)&  1631 &    925      &       1087       &    1460       &     1160
&     1081       \\
\hline
   Al(110)&  3393 &    925      &       1679       &    2870       &     1700
&     1090       \\
\hline
\end{tabular}
\end{center}
\label{tab2}
\end{table}


\newpage

\begin{thebibliography}{12}

\bibitem{S.E.1stPP-L.K.} N. D. Lang and W. Kohn, Phys. Rev. B 1 (1970) 4555.

\bibitem{W.F.1stPP-L.K.} N. D. Lang and W. Kohn, Phys. Rev. B 3 (1971) 1215.

\bibitem{S.E-2ndPP-R.D.} J. H. Rose and J. F. Dobson, Solid State Comm.
 37 (1981) 91; J. F. Dobson and J. H. Rose, J. Phys. C 15 (1982) 7429.

\bibitem{QUI-D.H.} J. F. Dobson and G. H. Harris, Phys. Rev. B 27 (1983)
6542.

\bibitem{2ndPPH-Barn.} R. N. Barnett, R. G. Barrera, C.L. Cleveland, and
Uzi Landman, Phys. Rev. B 28 (1983) 1667, and 28 (1983) 1685.

\bibitem{2ndPP.HA.-Egui.} A. G. Eguiluz, Phys. Rev. B 35 (1987) 5473.

\bibitem{SJM-P.T.S.} J. P. Perdew, H. Q. Tran, and E. D. Smith, Phys. Rev.
B 42 (1990) 11627 .

\bibitem{S.J.CAL.-FIO.} C. Fiolhais and J. P. Perdew, Phys. Rev. B 45
(1992) 6207.

\bibitem{2D.PP-Mon.P.} R. Monnier and J. P. Perdew, Phys. Rev. B 17 (1978)
2595; 22 (1980) 1134(E) .

\bibitem{SJM.REV.-Per.} J. P. Perdew, Prog. Surf. Sci. 48 (1995) 245.

\bibitem{SJM.F.D.-Kie.} A. Kiejna, Phys. Rev. B 47 (1993) 7361.

\bibitem{S.E.LMTO-Skriver} H. L. Skriver and N. M. Rosengaard, Phys. Rev. B
43 (1991) 9538, and 46 (1992) 7157.

\bibitem{comp.S.E.-Ing.} J. E. Inglesfield, in {Cohesion and Structure
of Surfaces}, F. R. De Boer and D.G. Pettifor (Eds.), Elsevier,
Amsterdam, 1995, Vol. 4.

\bibitem{QSE.-Sch} F. K. Schulte, Surf. Sci. 55 (1976) 427.


\bibitem{QSE-we} I. Sarria, C. Henriques, C. Fiolhais, and J. M. Pitarke,
Phys. Rev. B 62 (2000) 1699.

\bibitem{PP-Ash} N. W. Ashcroft, Phys. Lett. 23 (1966) 48.

\bibitem{PP.H-A.} V. Heine and D. Weaire, in { Solid State Physics},
 H. Ehrenreich, F. Seitz, and D. Turnbull (Eds.), Academic, New York,
1970, Vol. 24; I. Abarenkov and V. Heine, Philos. Mag. 12 (1965) 529;
V. Heine and I.Abarenkov, Philos. Mag. 9 (1964) 451.

\bibitem{PP.Evan.} C. Fiolhais, J. P. Perdew, S. Q. Armster, J. M.
MacLaren, and M. Brajczewska, Phys. Rev. B 51 (1995) 14001; 53 (1996) 13193
(E).

\bibitem{EC.bulk-Fern.} F. Nogueira, C. Fiolhais, and J. P. Perdew, Phys.
Rev. B 59 (1999) 2570.

\bibitem{Pit} J. M. Pitarke and A. Eguiluz, Phys. Rev. B 57 (1998) 6329.

\bibitem{XC-K.P.} S. Kurth and J. P. Perdew, Phys. Rev. B 59 (1999) 10461;
Z. Yan, J.P. Perdew, S. Kurth, C. Fiolhais, and L. Almeida, Phys. Rev. B 61
(2000)
2595.

\bibitem{EC.Clusters} F. Nogueira, C. Fiolhais, J. He, J. P. Perdew, and A.
Rubio, J. Phys: Cond. Matt. 8 (1996) 287.

\bibitem{EC.surf.clust} C. Fiolhais, F. Nogueira, and C. Henriques, Prog.
Surf. Sci. {53}, 315 (1996).

\bibitem{Li} U. Birkenheur, J.C. Boettger, and N. Rosch, Surf. Sci.
341 (1995) 103.

\bibitem{SAJM-Montag} B. Montag, P.-G. Reinhard, and J. Meyer, Z. Phys. D.
32 (1994) 125.

\bibitem{XC-P.W.} J. P. Perdew and Y. Wang, Phys. Rev. B 45 (1992) 13244.

\bibitem{ATOM} J. Sch\"ochlin, K.P. Bohnen, and K. M. Ho, Surf. Sci. 324
(1995) 113.


\end{thebibliography}
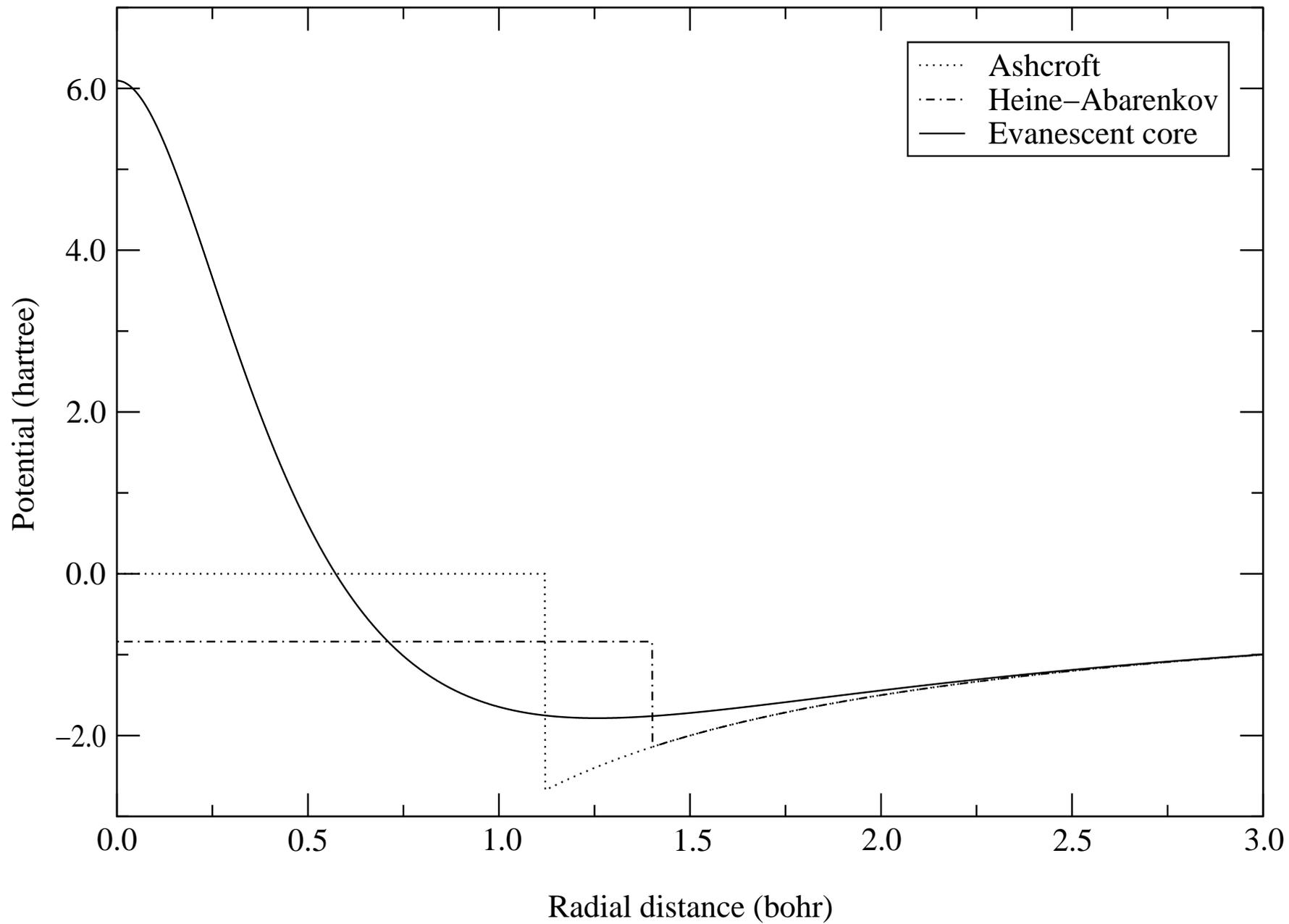
\end{document}